\documentclass[preprint2]{aastex}
\newcommand{\lapprox}{\mbox{$\stackrel{<}{_{\sim}}$}}
\newcommand{\gapprox}{\mbox{$\stackrel{>}{_{\sim}}$}}

\slugcomment{to appear in the Astrophysical Journal}

\shorttitle{C$^{18}$O in the Nucleus of IC 342}
\shortauthors{Meier \& Turner}

\begin{document}
\input{psfig.sty}

\title{Molecular Gas and Star Formation in the Nucleus of IC 342: 
C$^{18}$O and Millimeter Continuum Imaging}

\author{David S. Meier, \& Jean L. Turner} 
\affil{Department of Physics and Astronomy,
University of California, Los Angeles, CA 90095--1562
\\email: meierd@astro.ucla.edu; turner@astro.ucla.edu}

\begin{abstract}

We present high resolution maps ($\sim 2^{''}$) maps of the J=1-0 and
J=2-1 transitions of C$^{18}$O in the central $\sim 150$ pc of the
gas-rich nucleus of IC 342, made with the Owens Valley Millimeter
Array.  From the C$^{18}$O maps we are able to obtain the most
accurate map of N$_{H_{2}}$ to date for IC 342.  Due to their low
opacities, the transitions of C$^{18}$O give a more reliable estimate
of the true molecular gas column density distribution than the more
common $^{12}$CO and $^{13}$CO isotopomers.  The morphology of the
C$^{18}$O emission in the nucleus is a mini-spiral similar to that of
the main isotopomer, $^{12}$CO, except it is more symmetric, lacking
the enhancements to north.  We suggest that the asymmetries present in
$^{12}$CO images may reflect the viewing perspective of the starburst
region biased by the high optical depths of $^{12}$CO, rather than
true asymmetries in the amount of molecular gas present.  The giant
molecular clouds seen in C$^{18}$O appear to be nonspherical, probably
due to tidal arm shearing.  Column densities determined from C$^{18}$O
observations, 1.3 mm dust continuum, and the virial theorem indicate
that the standard Galactic conversion factor, X$_{CO}$, overestimates
the amount of molecular gas in the center of IC 342 by a factor of
$\sim$2-3 at the molecular cloud peaks, and by $>$3 in the diffuse gas
away from the starburst.  Revised molecular masses based on this
conversion factor imply that star formation efficiencies in the
starburst region are very high.  From the distribution of gas and star
formation it appears that the sites of star formation are dynamically
determined, rather than driven by density peaks.  Near the central
star-forming region, evidence is seen for chemical enrichment of
C$^{18}$O due to massive stars.
\end{abstract}
\keywords{galaxies:individual(IC342)---galaxies:ISM---galaxies:
nuclei---galaxies:starburst---radio lines:galaxies}

\section{Introduction}

Our current knowledge about molecular gas in galaxies is primarily
based on the bright, optically thick lines of $^{12}$CO, using the
Galactic standard conversion factor, $X_{CO}$. These assumptions,
while probably adequate when applied to the disks of large spiral
galaxies such as our own \citep{SRBY87, Sco87, WS90, NK95}, may not
apply in the unusual conditions present in nuclear starbursts.
Enhanced metallicities, densities and temperatures, large turbulent
linewidths, and intense radiations fields that are present in galactic
centers may change the value of $X_{CO}$ \citep[eg.,][]{MB88, E89,
Sak96}.  It has been suggested that the $X_{CO}$ overestimates gas
masses in our own Galactic Center by factors of 5 - 16 \citep{Set94,
DHWM98}, and in nearby starbursts \citep*[eg.,][]{SBMPN91, Xie94,
MHWCT96, BGDBHW97}.

Optically thin lines are promising probes of the molecular gas,
because they are able to ``sample'' all the column density and not
just conditions at cloud edges.  Unfortunately, because the abundances
of the isotopic species are much lower, the emission lines are
relatively weak.  Only recently has instrumentation reached the point
where interferometeric studies of isotopomers in other galaxies are
feasible.  High resolution studies of $^{13}$CO
\citetext{e.g. \citealp{Maff291}; Turner \& Hurt 1992, hereafter
\citealp*{TH92, PSS96, ARSS97, NGKGW98, MKVTK98, RD99}; Meier, Turner
\& Hurt 2000, hereafter MTH00}, indicate that even $^{13}$CO is
optically thick in the massive clouds of starburst nuclei \citep{WJ90,
PSS96, MTH00}.  Therefore, C$^{18}$O may be the best hope for
determining true molecular gas distributions of nuclear starburst
regions.

IC 342 provides an excellent galaxy for studying the relative
locations of different molecular gas tracers within the nucleus of
galaxies because it is the closest large spiral with active nuclear
star formation \citetext{1.8 Mpc, Table 1: \citealp{Mc89, MF92, KT93};
see \citealp{BM99} for a detailed discussion of the distance to IC
342}.  IC 342 is nearly face-on, with widespread resolvable molecular
line emission both in dense clouds and a diffuse medium
\citep[eg.,][]{Loet84, I90, D92, WITHL93}.

In this paper, the true distribution of the molecular gas in the
nucleus of IC 342 will be investigated.  Questions such as the
following will be addressed: what is the distribution of H$_{2}$?
What is $X_{CO}$?  Does $X_{CO}$ vary over the starburst region? and
does it correlate with the observed physical conditions of the
molecular gas?  With these goals in mind, we obtained $\sim2^{''}$
resolution Owens Valley Millimeter Interferometer images of
C$^{18}$O(2-1), C$^{18}$(1-0), 2.6 mm and 1.3 mm continuum of the
central region of IC 342, giving spatial resolutions of $\sim 15~-~
20$ pc.  The C$^{18}$O(2-1) was observed simultaneously with
C$^{18}$O(1-0) to allow determination of T$_{ex}$, which is necessary
for the determination of optically thin column densities. These
observations are combined with previous data of the J=2-1 and 1-0
transitions of $^{12}$CO and $^{13}$CO \citep*{TH92, THH93, MTH00} to
obtain a detailed and accurate description of the molecular gas column
density within the central 150 pc.

\section{Observations}

Aperture synthesis observations of both C$^{18}$O(1-0) (109.782 GHz)
and C$^{18}$O(2-1) (219.560 GHz) were made with the Owens Valley Radio
Observatory (OVRO) Millimeter Interferometer between 1997 October 22 and
1997 December 15 (Table 2).  The interferometer consists of six 10.4
meter antennas with cryogenically cooled SIS receivers \citep{OVRO91,
OVRO94}.  System temperatures (single sideband) ranged from 250 to 400
K at 2.6 mm and 450 - 1000 K at 1.3 mm.

The transitions were observed simultaneously making use of OVRO's
multiple-line spectrometer capabilities.  Therefore, both transitions
have consistent instrumental configurations, phase centers, and
weather.  The (2-1) data have the same (u,v) pattern but with twice
the scale as the (1-0) data.  A filterbank with 32 4 MHz channels was
used, giving velocity resolutions (bandwidths) of 10.9 km s$^{-1}$
(349 km s$^{-1}$) at 2.6 mm and 5.5 km s$^{-1}$ (175 km s$^{-1}$) at 1
mm.  IC 342's systemic velocity with respect to LSR is 35 km s$^{-1}$
and was centered at channel 16.5.  Two pointings with phase centers of
$\alpha_{1}(B1950) = 03^{h}41^{m}57.^{s}9$; $\delta_{1}(B1950) =
+67^{o}56^{'}29.^{''}0$ and $\alpha_{2}(B1950) =
03^{h}41^{m}57.^{s}0$; $\delta_{2}(B1950) = +67^{o}56^{'}26.^{''}0$
were observed to cover the entire nuclear mini-spiral.  Data were
calibrated using the MMA software package.  Calibrators 3C84 and
0224+671 were observed in 20 minute intervals for phase calibration.
Absolute flux calibration was done using Neptune and Uranus as primary
flux calibrators and 3C273 as a secondary flux calibrator for both
transitions.  Absolute flux calibrations are estimated to be good to
10 - 15\% for the 2.6 mm data and 20 - 25\% for the 1.3 mm data.

The dataset was mosaicked using the MIRIAD package.  The maps are
naturally weighted and primary-beam corrected.  Since the maps are
mosaics, noise levels vary across the maps.  Noise levels reported
throughout this paper are those measured from line free channels of
the spectral cube half-way between the map center and the FWHM points.
The true noise level is slightly lower ($\sim 10\%$) at the phase
centers and somewhat higher towards the edge of the primary beams.
All further data reduction was done with the NRAO AIPS package.  In
making the zeroth moment (integrated intensity) maps, only emission
greater than 1.2$\sigma$ was included.  Due to the fact that the
signal-to-noise (S/N) is lower in the C$^{18}$O(2-1) map, the choice
of 1.2$\sigma$ could, in principle, lower the integrated intensities
of the C$^{18}$O(2-1) map relative to the higher sensitivity
C$^{18}$O(1-0) map.  However, the ratio of the integrated intensity
maps made with no clipping is identical to those with clipping to
within 5\%, indicating this effect is not important.

The (u,v) coverage for these observations imply that emission on
scales larger than $\sim$36$^{''}$ (3 mm) or $\sim$18$^{''}$ (1.3 mm)
is resolved out.  To estimate the amount of extended flux the
interferometer resolves out in the C$^{18}$O(1-0) map, it was compared
with the single-dish spectra from the 30 m IRAM telescope \citep{E90}
at 21$^{''}$ resolution.  The integrated intensity of the OVRO map
when convolved to 21$^{''}$ is 3.6 $\rm K~km~s^{-1}$, which is
$\sim$67\% of the 4.8 $\rm K~km~s^{-1}$ detected by IRAM.  This is
consistent with the 70 - 75\% detected in the $^{12}$CO(1-0) map
\citep*{LTH94}, and not much lower than the 80\% estimated for the
$^{13}$CO transitions \citep{TH92, MTH00}.  A similar analysis yields
4.9 $\rm K~km~s^{-1}$ for C$^{18}$O(2-1) or $\sim$60\% of the
single-dish flux.  Therefore, the amount of flux resolved out in
$^{12}$CO(1-0), $^{13}$CO, C$^{18}$O are all similar and should not
strongly effect the inter-comparisons.

\section{Results}

\subsection{The C$^{18}$O Morphology}

The C$^{18}$O(1-0) and C$^{18}$O(2-1) channel maps are presented in
Figures 1-2.  Peak antenna temperatures obtained in the 4 MHz channels
are $\sim$ 0.75 K for C$^{18}$O(1-0) and 1.5 K for C$^{18}$O(2-1).
Emission is present in channels from V$_{LSR}$ = -14 km s$^{-1}$ to 73
km s$^{-1}$ with higher velocities in the northeast.  The line center
and velocity dispersions of both transitions of C$^{18}$O are
consistent with those of $^{13}$CO \citep{TH92}.  Integrated intensity
maps of the C$^{18}$O(1-0) and C$^{18}$O(2-1) transitions, along with
the $^{12}$CO(1-0) map \citep{LTH94} are plotted to the same scale in
Figure 3.  Integrated intensities of each transition are listed in
Table 3 for various locations across the nucleus
\citetext{nomenclature from \citealp{D92}, Figure 2 of
\citealp{MTH00}}.

The basic morphology of the nuclear mini-spiral seen in C$^{18}$O(1-0)
is similar to the $^{12}$CO(1-0) transition, but there are noticeable
differences.  Observations of $^{12}$CO and HCN show that there are
five major peaks of molecular gas in the nucleus \citep{I90, D92}.
GMC C, along the northern arm, is brightest in $^{12}$CO, GMC B is the
molecular cloud associated with the starburst, and GMC A is the
nearest to the dynamical center of IC 342.  GMCs D and E are more
distant GMCs along the northern and southern arms, respectively.  All
GMCs except GMC A are readily seen in C$^{18}$O(1-0).  GMC A is
substantially weaker in the 1-0 transition of C$^{18}$O relative to
the $^{12}$CO(1-0) transition.  However, in the 2-1 transition of
C$^{18}$O, GMC A is similar in relative strength.  The weakness of GMC
A in C$^{18}$O(1-0) is probably a temperature effect (\S3.2.1).

Another difference is that C$^{18}$O(1-0) distribution, like
$^{13}$CO(1-0) \citep{TH92, WITHL93}, appears more symmetric than
$^{12}$CO(1-0).  In $^{12}$CO(1-0), the northern arm is brighter than
the southern arm by about 50\% (the integrated flux of the entire
northern arm is 1200 Jy km s$^{-1}$, while the entire southern arm is
800 Jy km s$^{-1}$).  In C$^{18}$O(1-0) it is only $\sim$15\% percent
brighter (the integrated flux of the entire northern arm is
15.4$\pm$2.5 Jy km s$^{-1}$, as compared to 13.6$\pm$2 Jy km s$^{-1}$
for the southern arms).  C$^{18}$O(1-0) preferentially arises from the
outside (leading) edges of the arms, similar to the $^{13}$CO(1-0),
while $^{12}$CO(1-0) comes from broad arms with abundant diffuse
emission upstream from the arms.  The less abundant isotopomers do not
show the broad diffuse arms nearly as dramatically as the $^{12}$CO
\citep{WITHL93}.  Although the sensitivity of the C$^{18}$O(1-0) maps
are not extremely high, the high sensitivity $^{13}$CO(1-0) line
displays the same trend.  These differences between $^{12}$CO and the
rarer isotopes can be understood if the geometry and excitation of the
nuclear gas clouds are considered.

\subsubsection{The Geometry of the Starburst}

From Figure 4, it can be seen that the two molecular arms in the
nucleus of IC 342 progress inward and wrap around the central
starburst.  The arms appear to terminate in a partial ring surrounding
the burst.  If the two arms are due to a trailing density waves with
orbital motions predicted for barred spiral arms outside of the Inner
Lindblad Resonance (ILR) \citep*[eg.,][]{RHV79, A92}, then the
velocity pattern implies the northern arm is behind the starburst in
projection and the southern arm is in front.  Comparisons of H$\alpha$
to thermal radio continuum and observations of B$\gamma$/H$\alpha$
confirm that extinction towards the southern portion of the starburst
complex is substantially higher than towards the north \citep*[Figure
4; Figure 5 of][]{MTH00, BFG97}.

This geometry implies that from our perspective, we are seeing the
``starburst lit'' face of the molecular clouds when viewing the
northern arm (seeing the surface PDR regions), and the ``backside'' of
the molecular clouds for the southern arm (seeing the ambient
molecular gas).  For optically thick transitions such as those of
$^{12}$CO, which trace the cloud photospheres \citep{THH93}, the lines
should appear brighter along the northern arm relative to the southern
arm, hence are sensitive to temperature effects.  Whereas the
optically thin transitions will trace column density.  Therefore this
asymmetric pattern can manifest itself in optically thick transitions,
but not in transitions that are either inherently optically thin due
to opacity, or effectively thin due to large velocity gradients.

\subsubsection{Giant Molecular Clouds in IC 342}

Because of its low opacity and higher critical density, C$^{18}$O(2-1)
emission, like $^{13}$CO(2-1), is dominated by the dense giant
molecular clouds (GMCs) seen in HCN emission \citep{D92}.  The
C$^{18}$O(2-1) observations have the highest spatial resolution yet
obtained ($\sim$ 13 pc) in IC 342, sufficient to resolve the nuclear
GMCs into substructure.  Therefore, C$^{18}$O(2-1) is used to derive
the properties of the individual GMCs.  Each cloud, identified only as
a region of spatially and spectrally localized emission, not
necessarily a gravitationally bound entity, has been fit with an
elliptical Gaussian to determine its size, linewidth, location and
intensity.  Of the five GMCs identified by \citeauthor{D92}, only GMCs
A, B, C and E are fit since GMC D is at the edge of the C$^{18}$O(2-1)
primary beam.  The four major GMCs are resolved into eight components
(numbers are added to the letter designations in the case of
substructure).  The results of the fits are tabulated in Table 4.

All clouds are resolved (fitted sizes $>$1/2 the FWHM of the beamsize)
except cloud C3.  The GMCs are composed of several smaller clouds that
are $\sim 10\times30$ pc in size, with masses of about $ \rm \sim
10^{6} ~M_{\odot}$ and FWHM linewidths of 20 - 30 $\rm K~km~s^{-1}$.
The GMCs tend to be non-spherical with axial ratios $\sim$2 and
position angles generally oriented in the direction of the spiral
arms.  This is consistent with the presence of arm streaming motions
over the nuclear region \citep{D92, TH92}.  Beam averaged densities
for the molecular clouds that make up the GMCs are $\sim 10^{3.5}~\rm
cm^{-2}$ (\S 3.2.3).  These values are quite similar to those found
for the ``envelope'' of the Galactic Center GMC, Sgr B2
\citep[eg.,][]{GBMZHKSL93, HWMLDH95}.

\subsection{Comparing CO Isotopomers: Gas Excitation and Opacity in IC 342} 
\subsubsection{Excitation Temperatures from C$^{18}$O(2-1)/C$^{18}$O(1-0)}

Comparisons between rotational lines with different J states of
optically thin transitions allow for the mapping of gas excitation
temperatures (T$_{ex}$).  Generally, the main isotopomers of CO are
optically thick in galactic nuclei and as such are of little use in
constraining T$_{ex}$ \citep[eg.,][]{BC92}.  Moreover, gradients in
temperature over small scales appear to dominate the $^{12}$CO line
ratios rather than T$_{ex}$ \citep{THH93}.  Optically thin line
ratios, on the other hand, can be quite constraining, making the
C$^{18}$O(2-1)/C$^{18}$O(1-0) line ratio a sensitive measure of
T$_{ex}$ in IC 342.  The ratio of the C$^{18}$O(2-1) integrated
intensity to the C$^{18}$O(1-0) intensity, under the LTE assumption
is:
$$
{\int ^{18}T_{21}dv \over \int ^{18}T_{10}dv}~ \simeq
$$
$$
~{^{18}f_{21}(^{18}J_{21}(T_{ex}) - ~^{18}J_{21}(T_{cmb}))
(1-e^{-^{18}\tau_{21}})\over ^{18}f_{10}(^{18}J_{10}(T_{ex}) -~ 
^{18}J_{10}(T_{cmb}))(1-e^{-^{18}\tau_{10}})},
$$
where $J_{\nu}(T_{ex})~=~(h\nu/k)/(exp\{h\nu/kT_{ex}\}-1)$, $\tau$
and $f$ are the optical depth and filling factor of each transition,
respectively.  

In Figure 5a, the C$^{18}$O(2-1)/C$^{18}$O(1-0) line ratio map is
presented for the nucleus of IC 342.  Ratios of integrated intensities
are used instead of the peak main-beam temperatures to increase the
S/N.  This requires that the C$^{18}$O(2-1) and C$^{18}$O(1-0) line
profiles be the same.  This is a good approximation (compare Figure 1
and 2).  In generating the ratio maps, the integrated intensity maps
for C$^{18}$O(2-1) and C$^{18}$O(1-0) were convolved to matching
beamsizes, divided, and blanked anywhere where emission was $<~ 3
\sigma$ in either map.  Errors in the ratio maps are conservatively
estimated to be $\leq$35\% in magnitude and $\leq 1^{''}$ in position.

Aside from the two peaks seen in the ratio map, the
C$^{18}$O(2-1)/C$^{18}$O(1-0) ratio map has relatively constant values
of 0.7 - 1.3 over much of the central mapped region, with lower values
preferentially found in the south.  The highest
C$^{18}$O(2-1)/C$^{18}$O(1-0) peak, with a value of 2.2, is found at
the starburst region.  The heating source for the warm molecular gas
at this location is probably the massive stars of the starburst.  The
second ratio peak is at GMC A.  GMC A is unusual in that it is not
near either of the major radio continuum sources (\S 3.3), which
suggests that the heating of GMC A is not due to star formation.  Some
dynamical heating mechanism may be at work here, such as shock heating
due to large scale bar-driven gas flows, which is consistent with the
fact that it is the GMC closest to the dynamical center of the galaxy,
and to the ring of hot shocked gas seen in H$_{2}$ \citep{BFG97}.

A map of T$_{ex}$ for the nuclear region derived using the
C$^{18}$(2-1)/C$^{18}$(1-0) ratio is displayed in Figure 5b.  T$_{ex}$
ranges from 7 - 10 K over much of the mapped region.  The T$_{ex}$ of
the gas near the starburst region is 19$\pm$8 K.  The secondary
T$_{ex}$ peak at GMC A has T$_{ex}$ = 13$\pm$4 K.  Excitation
temperatures derived from the C$^{18}$O transitions are lower than
those estimated from $^{13}$CO \citep{MTH00}.  We estimate filling
factors using $^{18}f_{10}~ \simeq ~ $T$_{mb}/$T$_{ex}\tau_{10}$.  For
T$_{ex}~\simeq ~ 10$ K and $\tau_{10}~\sim~ 0.1 - 0.2$ (\S 3.2.4), the
implied filling factors are $^{18}f_{10}~\sim~ 0.5 - 1$, which given
the different beamsizes, are consistent with those found for
$^{13}$CO.  This implies that the C$^{18}$O(2-1) emission detected
arises predominately from the same molecular gas component as the
$^{13}$CO(2-1) emission \citep{MTH00}.

\subsubsection{Molecular Gas Opacities and Abundances: 
$^{12}$CO(1-0)/C$^{18}$O(1-0) \& $^{13}$CO(1-0)/C$^{18}$O(1-0)}

Comparison of the abundant and optically thick $^{12}$CO line with its
less abundant and thinner cousins, $^{13}$CO and C$^{18}$O, can be
used to trace optical depths and place limits on isotopic abundances.
The isotopic line ratios are displayed in Figure 6a-c and Table 3.
All of the isotopic line ratios discussed in this paper are between
the (1-0) transitions of each isotopomer and so we suppress the (1-0)
notation for the remainder of the paper.  The $^{12}$CO/C$^{18}$O line
ratio ranges from 11 - $\gapprox$110 over the mapped region.  Regions
that are blanked due to low S/N tend to correspond to high
$^{12}$CO/C$^{18}$O ratios since the weakness of the C$^{18}$O(1-0)
line is the limiting factor.  At the sites of the GMCs,
$^{12}$CO/C$^{18}$O ratios of 30 - 60 are found, while values for
$^{13}$CO/C$^{18}$O are 5-6.  Typical Galactic Center values for the
$^{12}$CO/C$^{18}$O and $^{13}$CO/C$^{18}$O are $\sim 60 - 80$ and
$\sim ~10$, respectively \citep[eg.,][]{OHSTM98, DHWM98}.  As in the
Galactic Center \citep{DHWM98}, variations in this ratio in the
nucleus of IC 342 are substantial over small ($\sim$30 pc) spatial
scales.  The highest column density regions at peaks B, C, and E tend
to have the lowest isotopic ratios.  This is expected since these are
presumably the optically thickest regions.

The Central Trough region has surprisingly low values of
$^{12}$CO/C$^{18}$O and $^{13}$CO/C$^{18}$O line ratios.  The Central
Trough region, which is a site of weak emission in all observed
transitions, is the only firmly established region of optically thin
$^{13}$CO gas over the central distribution \citep{MTH00}.  As such,
it is expected to have high isotopic ratios and give the actual
$^{13}$CO/C$^{18}$O isotopic abundance ratios.  Surprisingly, both
isotopic ratios, $^{12}$CO/C$^{18}$O = 24 and $^{13}$CO/C$^{18}$O =
3.2, are $lower$ here than the average across the rest of the spiral
arms (C$^{18}$O overluminous).  Along the spiral arms, the $^{12}$CO
tends to dominate C$^{18}$O (highest ratios) on the upstream sides of
the arms.  C$^{18}$O becomes relatively more prominent as one moves
towards the leading edge of the spiral arms (low ratios).  The same
effect is seen in $^{12}$CO/$^{13}$CO observations with higher
sensitivity to extended structure \citep{WITHL93}.

Much of the structure seen in the isotopic ratio maps appear to be due
to excitation effects.  The peak in the $^{12}$CO/C$^{18}$O ratio map
at GMC A and towards the starburst near GMC B result from depopulation
of the 1-0 level in the optically thin C$^{18}$O due to the increased
T$_{ex}$, as confirmed by the high C$^{18}$O(2-1)/C$^{18}$O(1-0)
ratios.  The trend of decreasing ratio counter-clockwise across the
arms likely results from changes in density across the spiral arm due
to the density wave \citep[][]{WITHL93}.  The variation in excitation
demonstrates the importance of observing higher J transitions when
dealing with optically thin transitions.  If one relies, as is often
the case for extragalactic studies, on just the observations of the
(1-0) transition of CO and its isotopomers, it is possible that
variations in excitation can be mistaken for variations in
morphological structure or column density.

A lower limit to the true $^{12}$CO/C$^{18}$O abundance ratio can be
estimated from the observed ratio.  In the limit of both transitions
being optically thin (LTE), the isotopic line ratio will reflect the
true isotopic abundance ratio.  Since $^{12}$CO(1-0) is not optically
thin, the ratio is a lower limit.  For conservative lower limits to
[$^{12}$C/$^{13}$C] and [$^{16}$O/$^{18}$O], we take the highest
(trustworthy) values of the ratio.  The highest observed value of the
$^{12}$CO/C$^{18}$O line ratio is $\simeq$110.  The
[$^{16}$O/$^{18}$O] abundance ratio must be higher than this.  The
peak values of the $^{13}$CO/C$^{18}$O line ratio are about 11.
However, both of these values are overestimates because of the non-LTE
effects discussed in \S3.2.4. When we have corrected for the non-LTE
effects, we estimate [$^{16}$O/$^{18}$O] $>$ 110 and
[$^{12}$C/$^{13}$C] $>$ 20.  Though not well constrained, the values
are consistent with those assumed as well as the (also weakly
constrained) values previously estimated for IC 342 based on other
molecular species \citep{HCMW98}.

The low isotopic ratios seen towards the Central Trough appear to
require truly anomalous abundances.  The low ratios in this region
imply that C$^{18}$O is overabundant relative to $^{12}$CO and
$^{13}$CO.  Neither isotope-selective photodissociation nor
isotope-selective fractionation can be the explanation because in both
cases the abundance of $^{12}$CO and $^{13}$CO are predicted to be
larger relative to C$^{18}$O, not smaller \citep{LGCW80, LGFA84,
VB88}.  This makes the most likely possibility for the higher
abundance of C$^{18}$O enrichment from $^{18}$O rich, massive star
ejecta.  A complete knowledge of the cooking sites of $^{18}$O has yet
to be achieved, but it appears that $^{18}$O is a secondary product of
massive stars \citep*[see eg.,][]{WM92, HM93, WR94, PAA96}.  If the
central starburst is biased towards massive stars, a relative increase
in the abundance of $^{18}$O (and hence C$^{18}$O) is possible.  Low
$^{12}$CO/C$^{18}$O values have been found before in other starburst
galaxies such as NGC 253, M 82, NGC 4945 as well as the AGN NGC 1068
\citep[eg.,][]{HM93, MHWCT96, PSS96}.  Using a chemical evolution
model, \citet{HM93} have modeled the isotopic abundance ratio for a
starburst typical of a galactic nuclei like M 82, NGC 253 and possibly
IC 342.  They find that the [$^{16}$O/$^{18}$O] can drop as low as
$\sim$150.  IC 342 also has a rather high
C$^{18}$O(1-0)/C$^{17}$O(1-0) line ratio of 6.5, consistent with
enhanced $^{18}$O \citep[L. Sage, from][]{HWLCM94}.  Therefore, for
the Central Trough region, it is possible that the [C$^{18}$O/H$_{2}$]
abundance may be higher than has been assumed.

\subsubsection{Molecular Gas Density: C$^{18}$O LVG modeling}

A sample of Large Velocity Gradient (LVG) radiative transfer models
were run to relate the observed intensities and line ratios to the
physical properties of the cloud, $n_{H_{2}}$ and T$_{k}$
\citep*[eg.,][]{GK74, SS74, DCD75}.  LVG models are probably not
appropriate for all lines in these tidally disturbed clouds,
\citep{THH93, MTH00}, but are nonetheless instructive, particularly
when the transitions get optically thinner, as is the case for
$^{13}$CO and C$^{18}$O.  The LVG model used is the same as discussed
in \citet{MTH00}, except a [C$^{18}$O/H$_{2}$] abundance of $3.4
\times 10^{-7}$ is used (\S 4.1).  The ratio of cloud linewidth to
cloud size is used to constrain the velocity gradient for the models
(Table 4).  Typical values of the velocity gradients for the GMCs are
$\sim ~1~-~5 \rm ~km~s^{-1}pc^{-1}$.  In Figure 7, we display two
models with velocity gradients of $\sim ~1$ and 5 $\rm
km~s^{-1}pc^{-1}$, for both $^{13}$CO and C$^{18}$O \citep[see][for
details of the $^{13}$CO models]{MTH00}.

Since C$^{18}$O is optically thinner than $^{13}$CO, it is less
affected by radiative trapping and thermalizes at a slightly higher
density than the corresponding lines of $^{13}$CO.  LVG modeling of
C$^{18}$O may be capable of constraining the densities, whereas
$^{13}$CO has only provided lower limits to the density.  Best fit
solutions obtained from the C$^{18}$O temperatures and line ratios are
indeed close to the lower end of the solution range found in
$^{13}$CO, and well constrained.  The C$^{18}$O observations imply
that most of the central GMCs have $<n_{H_{2}}> ~\simeq ~
10^{3.0-3.5}~ \rm cm^{-3}$ and T$_{k} ~\simeq ~ 10 - 40$ K.  These
moderate densities come directly from the fact that T$_{ex}$ derived
from C$^{18}$O are lower than those derived from $^{13}$CO.  These
densities are slightly smaller than those implied from the HCN
observations \citep{D92}, probably indicating that there is some
further clumping that is not resolved.  Closer inspection shows that
on average the C$^{18}$O solutions for GMC A, B and C imply slightly
warmer, lower density gas than does $^{13}$CO.  This is probably
caused by the fact that $^{13}$CO and C$^{18}$O were assumed to have
the same filling factor \citep[0.25 over a $4.^{''}8\times 3.^{''}9$
beam;][]{MTH00}.  Since the critical density is higher for C$^{18}$O,
one might expect cloud sizes to be slightly smaller for C$^{18}$O than
for $^{13}$CO.  If the relative filling factor for the C$^{18}$O
emitting gas is lowered, the models become consistent for all
locations.  In every case, the decrease in filling factor required for
agreement between $^{13}$CO and C$^{18}$O are less than a factor of
two.  It is interesting to compare the average densities derived from
the LVG analysis with what is predicted from high resolution
C$^{18}$O(2-1) and the virial theorem (\S 4.1).  The densities
estimated from both methods are not too different, with the values
derived from the virial theorem tending to be about 1.5 - 3 times
larger than the LVG models.

\subsubsection{Isotopic Line Ratios: Complications and Important 
Considerations}

These multi-line, multi-transition data on IC 342 provide a wealth of
information on the complex interactions between opacity and
temperature for molecular clouds in starburst environments.  Here we
consider what these multi-transition studies as a whole indicate for
the opacities of the CO lines.  The magnitude and effect of several
different non-LTE effects are estimated for the nucleus of IC 342.

For IC 342, the T$_{ex}$ and filling factors derived from the
$^{12}$CO, $^{13}$CO and C$^{18}$O lines can depend on the transition
considered.  A single T$_{ex}$ or filling factor cannot be assumed for
all the lines.  This complicates the determination of opacities and
abundances from the line ratios.  Since $^{13}$CO and C$^{18}$O have
lower opacities, their effective critical densities are higher than
for $^{12}$CO.  They are more confined to the cloud cores and will
have smaller relative filling factors \citep{WJ90, MTH00}.  For the
externally heated clouds in IC 342, the $^{12}$CO transitions will
also have higher T$_{ex}$ than the optically thin species
\citep{THH93, MTH00}.  Therefore the isotopic ratios between $^{12}$CO
and other isotopomers assuming LTE will underestimate the line
opacities.

It is difficult to estimate the degree to which differing filling
factors impact the line ratios independent of the opacity.  However,
differences in the T$_{ex}$ can be estimated since T$_{ex}$ have been
separately derived for $^{12}$CO, $^{13}$CO and C$^{18}$O \citep[][\S
3.2.1]{THH93, MTH00}.  If the peak brightness temperature of the
$^{12}$CO(2-1) transition is taken as a lower limit to the
$^{12}$T$_{ex}$, the $^{13}$CO(2-1)/$^{13}$CO(1-0) line ratio under
the optically thin assumption to estimate the $^{13}$T$_{ex}$, and
similarily the C$^{18}$O(2-1)/C$^{18}$O(1-0) line ratio for the
$^{18}$T$_{ex}$, then the magnitude of the temperature differences can
be estimated.  Over the central GMCs, typical derived temperatures are
$^{12}$T$_{ex}\gapprox~ 26$ K, $^{13}$T$_{ex}\simeq 20$ K and
$^{18}$T$_{ex}\simeq 10$ K. This implies that the line ratios
involving C$^{18}$O are $\sim$2 - 3 times larger than expected
assuming the same excitation temperature applies for all transitions.
Taking into account these corrections, the isotopic line ratios imply
$^{13}\tau_{10} \sim 0.7 - 3$ ($^{18}\tau_{10} \sim 0.1 - 0.5$).
These results are consistent with $^{13}\tau_{10} ~\gapprox~ 1$ as was
previously found from the $^{13}$CO line ratios \citep{MTH00}.  When
one compares isotopic line ratios in IC 342 to find abundances and
opacities, the derived values are underestimates because the optically
thinner transitions have lower T$_{ex}$ than the thicker transitions.
We expect that this variation in T$_{ex}$ for CO lines of different
opacities may be common in other starburst nuclei, and should be taken
into account when making estimates of opacities and abundances.

\subsection{Millimeter Continuum Morphology}

Continuum emission was detected at 2.6 mm and 1.3 mm in IC 342, in a 1
GHz continuum channel.  The 2.6 mm and 1.3 mm continuum maps,
convolved to the same beamsize, are displayed in Figures 8a-b.  The 2.6
mm continuum emission shows a twin-peaked structure similar in
morphology to the 2 cm and 3.4 mm continuum emission \citep{TH83,
D92}.  The two strongest sites of millimeter continuum are coincident
with the nuclear starburst (near GMC B) and a secondary cluster
centered on GMC C, respectively.  The peak intensity of 2.6 mm
continuum is 4.4 mJy bm$^{-1}$.

Figures 8c-d display the spectral index between 2 cm/2.6 mm and 2.6 mm/1
mm over the central region.  The 2 cm to 2.6 mm spectral index ranges
from $\alpha^{2}_{2.6} ~\simeq~-0.4$ to $0.3$ ($S_{\nu}\propto
\nu^{\alpha}$) over the map, similar to what is found for the spectral
index between 6 cm and 2 cm \citep{TH83}.  The generally flat spectrum
over the nuclear region is consistent with thermal bremsstrahlung
emission.  Lower values of $\alpha_{3}^{2}$ ($\sim -0.4$) towards the
northwest side of the central ring imply some contamination by
non-thermal synchrotron emission at 2 cm in this region.  Towards GMC
C, $\alpha_{2.6}^{2}$ is $\sim$+0.2 to +0.3.  Assuming all the 2.6 mm
continuum emission is from bremsstrahlung, the implied number of
ionizing photons, $N_{uv}$, present at each peak is $\simeq ~1.8\times
10^{51}~s^{-1}$, or about 80 O6 stars \citep*[assuming an T$_{e}$ =
8000 K; eg.][]{MH67, VGS96}.  The total flux in the mapped region is
S$_{2.6mm}$ = 26 mJy, implying that the entire central star-forming
region ($\sim$ 75 pc) has $N_{uv}~\simeq ~10^{52}~s^{-1}$
\citep[$\sim$ 460 O6 stars; consistent with that estimated
by][]{BGMNSWW80, TH83, D92}.

1.3 mm continuum emission is detected only from the molecular column
density peaks as traced by C$^{18}$O because the S/N is lower at this
frequency.  Peak 1.3 mm intensities are 8 mJy bm$^{-1}$ ($\sim 4\sigma$)
and 12 mJy bm$^{-1}$ ($\sim 6\sigma$) for GMC B and C, respectively.
Both 1.3 mm continuum sources have rising spectral indices of
$\alpha_{2.6}^{1.3}\sim$+2.5 between 2.6 mm and 1.3 mm.  Rising spectral
indices at 1.3 mm are seen tentatively towards GMC A as well (Figure
8d).  The 1.3 mm continuum appears to be largely thermal dust emission.
In fact for GMC C, the slightly rising spectral index between 2 cm and
2.6 mm indicates that a small amount of dust emission may even
contribute to 2.6 mm continuum at this location ($\lapprox$ 1 mJy
bm$^{-1}$).  We discuss the dust masses implied by the 1.3 mm continuum
in the following section.

\section{Discussion}

In order to study the relation of galactic dynamics, molecular gas,
and star formation, one needs accurate surface density maps of
molecular gas.  Estimates of column densities made using $^{12}$CO and
a conversion factor, $X_{CO}$ are commonplace, but require X$_{CO}$ to
be valid, which is not yet established.  Comparisons between masses
estimated using various methods need to be made in order to
investigate the validity of $X_{CO}$ in starburst environments.  Given
the wealth of data available for IC 342, and in particular the
valuable C$^{18}$O and 1.3 mm maps, accurate values of molecular
column density can be obtained.

\subsection{H$_{2}$ Gas Column Densities: Comparing C$^{18}$O, 
$X_{CO}$, Virial Masses and Dust}

Since C$^{18}$O is optically thin, H$_{2}$ column densities can be
obtained from the total C$^{18}$O(1-0) intensity:
$$
N(H_{2})_{C18O}=2.42\times 10^{14}~cm^{-2}~
{[~H_{2}~] \over [C^{18}O]}~
$$
$$
\times {e^{{5.27 \over T_{ex}}} \over 
(e^{{5.27 \over T_{ex}}} - 1)}~ I_{C18O}~(K~ km~ s^{-1}).
$$
We adopt an abundance ratio of [H$_{2}$]/[C$^{18}$O]=$2.94\times
10^{6}$ or [C$^{16}$O]/[C$^{18}$O]=250 \citep*{F82, HWLCM94, WR94}.
Figures 4 and 5c shows a map of $N_{H_{2}}$ calculated from optically
thin C$^{18}$O.  Within the 2-1 primary beam, the derived T$_{ex}$
obtained from the C$^{18}$O(2-1)/C$^{18}$O(1-0) line ratios are used
in calculating the column density.  Since T$_{ex}$ was determined to
be $\sim 7-10$ K (\S 3.2.1), T$_{ex}$ = 10 K is assumed when deriving
column densities for regions outside the C$^{18}$O(2-1) primary beam.
Column densities measured in the central region of IC 342 range from
$<0.25 - 4.8 \times 10^{22}~\rm cm^{-2}$ (corrected for resolved out
flux assuming a uniform distribution).  These values obtained for
optically thin C$^{18}$O can be compared with column densities
estimated using the a Galactic disk $X_{CO}$ of $\simeq~2 \times
10^{20} ~\rm cm^{-2}~(K~km~s^{-1})^{-1}$ \citep{Strong88, Het97}.
Using $X_{CO}$, we obtain N$_{H_{2}}$ column densities ranging from 2
- 12 $\times 10^{22} ~\rm cm^{-2}$ over the $\sim$ 25 pc beamsizes, a
factor of 3 to 13 higher than the column densities based on C$^{18}$O.

Two additional column density estimates are considered.  First, since
the GMCs in the center of IC 342 are resolved, the total molecular
mass can be estimated from the virial theorem.  For molecular clouds
with density profiles given by $\rho \propto R^{-\gamma}$, the virial
mass is given by \citep*[eg.,][]{MRW88}:
$$
M_{v}~=~ 126 \left(\frac{5 - 2 \gamma}{3 - \gamma} \right)
\left(\frac{\Delta v_{1/2}}{km~s^{-1}} \right)^{2}
\left(\frac{R}{pc} \right),
$$
where R is the radius of the cloud, and $\Delta v_{1/2}$ is the FWHM
of the linewidth.  The radius of the clouds is assumed to be
1.4$R_{1/2}$, or 0.7$\sqrt{ab}$ \citep*[eg.,][]{WS90, GBL97}, where
$a$ and $b$ are the FWHM fitted sizes of the major and minor axes.  A
$\gamma =1$ is assumed for the GMCs, consistent with what is found for
Galactic clouds \citep[eg.,][]{CBD85, SYCSW87, MRW88}.  However, the derived
masses are only weakly dependent on the choice of $\gamma$.  The
masses will change from their $\gamma=1$ values by less than a factor
of 50\% for any $\gamma$ between 0 - 2.  Virial cloud masses for
$\gamma = 1$ are displayed in Table 4.

Gas masses may also be estimated from optically thin thermal dust
continuum emission.  The 2.6 mm and 1.3 mm continuum maps are used to
measure the amount of dust.  Assuming the clouds are entirely
molecular and have a constant gas to dust ratio of 100 by mass, the
gas mass is related to the 1.3 mm dust continuum flux by
\citep[eg.,][]{H83}:
$$
M_{gas}(1.3~ mm)~=~310~ M_{\odot}
\left(\frac{S_{1mm}}{mJy} \right)
\left(\frac{D}{Mpc} \right)^{2}
$$
$$
\times 
\left(\frac{\kappa_{\nu}}{cm^{2}~g^{-1}} \right)^{-1}
\left(e^{\frac{10.56}{T_{d}}}-1\right),
$$
where $\kappa_{\nu}$ is the dust absorption coefficient at this
frequency, $S_{1mm}$ is the 1.3 mm flux, D is the distance and T$_{d}$
is the dust temperature.  The dust opacity, $\kappa_{\nu}$, at 1.3 mm
is taken to be $3.1 \times 10^{-3}~\rm cm^{2}~g^{-1}$, but is
uncertain by at least a factor of four \citep{PHBSRF94}.  A T$_{d}=42$
K is assumed based on the FIR colors \citep{BGMNSWW80, RH84}.  The
dust temperature applicable to the 1.3 mm observations could be lower
than this if a cool dust component missed by the FIR observations
exists.  Since dust emission is only detected at the two sites with
major star formation, we do not expect that a substantial cool dust
component exists.  Therefore, uncertainties in the derived gas mass
are dominated by the uncertainty in opacity..

The amount of dust emission is estimated by assuming the 2.6 mm
continuum emission is entirely thermal free-free emission.  The nearly
identical morphology and flat spectral index between 2.6 mm and 2 cm
makes this a reasonable assumption \citep{TH83}.  (In the youngest
star-forming regions, there are regions of compact optically thick HII
regions that do have rising spectral indices, [eg., \citealt*{BTHLK96,
THB98, BTK00}], but even these regions will be optically thin by 3
mm.)  Thermal free-free emission at 1.3 mm is removed by extrapolating
the 2.6 mm map to 1.3 mm using $\alpha~=~-0.1$.  The remaining flux of
$\sim$8 mJy bm$^{-1}$ at GMC C and $\sim$4 mJy bm$^{-1}$ at GMC B is
estimated to be dust emission.  Beam averaged column densities based
on dust emission are shown in Table 5.  The gas masses implied by the
dust are $ \rm 7.4\times 10^{5}~M_{\odot}$ and $\rm 3.6\times
10^{5}~M_{\odot}$ (or $\rm 7.7 \times 10^{22}~\rm cm^{-2}$ and $3.8
\times 10^{22}~\rm cm^{-2}$), for the two clouds detected at 1.3 mm,
respectively.  Uncertainties are at least a factor of 2 - 4, dominated
by uncertainties in $\kappa_{\nu}$.  Given the large uncertainties,
dust masses should be treated as indicative only.  Dust emission in IC
342 will be discussed further in a later paper (Meier et al., in
preparation).

\subsection{The Validity of $X_{CO}$ in IC 342's Nuclear GMCs}

The different methods of obtaining M$_{H_{2}}$ and N$_{H_{2}}$ can now
be compared to determine the validity of the $X_{CO}$ in the nucleus
of IC 342.  Column densities derived for the GMCs in IC 342 from the
three different methods described above are listed in Table 5.  At all
locations, the lowest column densities are estimated from the
C$^{18}$O data.  The C$^{18}$O column densities at the GMCs are 3 - 4
times lower than those found using the Galactic $X_{CO}$.  Column
densities based on dust emission are slightly higher but consistent
with those estimated from C$^{18}$O.  Column densities based on the
virial theorem tend to be the largest of the three methods and are
approximately equal to the values obtained using the Galactic
$X_{CO}$.

The dominant uncertainties in the column density estimates based on
the C$^{18}$O data are the [C$^{18}$O/H$_{2}$] abundance ratio and
differing source sizes.  A Galactic Center value of
[C$^{18}$O/H$_{2}$] has been assumed.  The isotopic line ratios, while
not strongly constrained, indicate that this value is reasonable (\S
3.2.2).  For the C$^{18}$O column density to be consistent with
$X_{CO}$ value, an abundance of [$^{16}$O/$^{18}$O] $\simeq$ 1000
would be required.  This value is larger than the highest ISM values
seen in the Galaxy ($\sim$600) and at least a factor of four larger
than is typically seen in galactic nuclei \citep[eg.,][]{HM93, WR94}.
If anything, the nucleus of IC 342 could have a [C$^{18}$O/H$_{2}$]
abundance that is higher than the $3.4\times 10^{-7}$ assumed, since
nuclear processing in massive stars tend to raise the
[C$^{18}$O/H$_{2}$] abundance rather than lower it (\S 3.2.2).  From
an abundance standpoint, we expect that the C$^{18}$O values listed in
Table 5 do not underestimate the molecular gas by a large amount.
From a photochemical standpoint, it is possible, given isotope
selective photodissociation and gas density variations that C$^{18}$O
emission will remain undetected in regions with detectable $^{12}$CO.
In IC 342, the cloud sizes do appear to be smaller in C$^{18}$O than
in $^{12}$CO \citep[\S 3.2.3;][]{MTH00}, so the column densities
estimated by C$^{18}$O cloud be slight underestimates.  Gas excitation
can also influence the optically thin estimates, although our dual
line observations mostly correct for excitation.

Column densities estimated from the virial theorem are very likely
overestimates of the true molecular gas column density.  The clouds
tend to be somewhat filamentary in appearance, consistent with tidal
shear or arm streaming motions \citep{TH92, D92}.  The linewidths of
the GMCs are probably larger than virial and the clouds may not be
bound.  In addition, linewidths in this region are slightly
overestimated simply by the finite resolution of the telescope in the
steep part of the rotation curve.  Not surprisingly, therefore, the
LVG models predict average densities a factor of 1.5 - 3 times lower
than the virial prediction.

While the dust masses are more uncertain due to assumptions of T$_{d}$
and opacity compared to the C$^{18}$O mass, they provide independent
indications for a lower $X_{CO}$ in the center of IC 342.  First, the
conversion factor estimated from dust continuum emission for GMCs B
and C, are intermediate between the Galactic $X_{CO}$ value and the
C$^{18}$O estimates.  Second, using near-infrared (NIR) spectroscopy
to measure the NIR extinction, \citet{BFG97} obtain A$_{v}~ \simeq
25$.  From the geometry of the starburst (\S 3.1.1), it appears that
GMC B is in front of the starburst.  The NIR extinction then implies a
column density of $\sim 2\times 10^{22}~\rm cm^{-2}$ towards the
starburst region (using the Galactic conversion between column density
and extinction).  This value is roughly consistent with the optically
thin C$^{18}$O value at this location.  If some of the gas is behind
the starburst then this is an underestimate.  But even if the
starburst is in the middle of the cloud, the observed value is still
below that estimated from $X_{CO}$.

{\it Considering all the different methods, and their possible biases,
it is concluded that $X_{CO}(IC ~ 342)$ $\simeq ~ 0.7 - 1 \times
10^{20}~\rm cm^{-2}~(K~km~s^{-1})^{-1}$ for the inner $\sim$75 pc of
IC 342, or about 1/2 - 1/3 the value of $X_{CO}$ for the Galaxy.}
While this value is not as low on average as is inferred for the
Galactic Center, it is closer to what is found towards the GMCs Sgr A
and Sgr B2 in the Galactic Center, based on C$^{18}$O \citep{DHWM98}.

\subsection{Variation of the $X_{CO}$ Across the Nuclear Mini-spiral}

Of equal importance to the mean $X_{CO}$ in the center of IC 342 is
its spatial variation and the general applicability of a constant
$X_{CO}$ over the nuclear region.  As noted above, there are some
differences between the morphology of the C$^{18}$O(1-0) and
$^{12}$CO(1-0) emission.  C$^{18}$O in the outer arm regions is
fainter than predicted using even the modified $X_{CO}$ suitable for
the central GMCs, implying $X_{CO}(IC ~342)$ is even lower in the
diffuse gas.  Plotted in Figure 9 (Table 5) is the estimated variation
in $X_{CO}(IC ~342)$ obtained from C$^{18}$O with distance from the
dynamical center for regions with detected C$^{18}$O.  $X_{CO}(IC
~342)$ decreases by $\sim$ 2 - 3 from the central value beyond the
inner $15^{''}$.  For the diffuse gas dehind the density wave,
C$^{18}$O(1-0) emission is not detected at all and $X_{CO}$ may be
even lower.

Since the C$^{18}$O(2-1) observations do not cover this region, we
have assumed T$_{ex}$ = 10 K, consistent with the average seen over
the mapped region.  Densities in the diffuse gas are probably lower
than towards the GMCs, so it is possible that T$_{ex}$ for C$^{18}$O
could be slightly lower than this due to its higher critical density.
Column densities estimated from the optically thin C$^{18}$O predict
$lower$ column densities for cooler temperatures (even larger
difference in $X_{CO}$).  Moreover, derived N(H$_{2}$) are only weakly
sensitive to T$_{ex}$ for such low temperatures (Eq.(2)).  {\it
Therefore, the possibility that C$^{18}$O is more strongly subthermal
in these regions cannot be the explanation for the differences derived
in the $X_{CO}$}.  This indicates that the widespread, diffuse gas
seen over much of the mapped region is substantially brighter in
$^{12}$CO than would be expected based on the C$^{18}$O data.  Even
higher values are found in the north, consistent with $^{12}$CO being
especially bright in the north (\S 3.1.1).

\citet{DHWM98} find that diffuse gas in the Galactic Center also has a
much smaller $X_{CO}$ than is estimated for Sgr A and Sgr B2, and that
$X_{CO}$ is not valid across the entire region.  Because they did not
have C$^{18}$O(2-1) observations to go along with C$^{18}$O(1-0) and
instead relied upon global LVG models to predict the C$^{18}$O
excitation, some of the structure they find probably reflects
variations in excitation.  Such variations are important in IC 342 (\S
3.2.2).  Hence, simply using a ``modified'' $X_{CO}$ is only good
enough to characterize the molecular gas distribution in nuclei to a
factor of $\sim$2, with the amount of diffuse gas being overestimated.
Indications of similar variations in the estimated $X_{CO}$ with
distance from the center are also seen in M 82, and Maffei 2
\citep[][Meier, Hurt \& Turner, in preparation]{SBMPN91}.

\subsection{An Unusually High Star Formation Efficiency: Dynamically 
Determined Star Formation in IC 342?}

The revised H$_{2}$ column densities have implications for the nature
and efficiency of star formation in IC 342.  Since $^{12}$CO appears
to slightly overestimate the molecular column densities, previously
derived molecular mass fractions for the central region of IC 342 need
to be revised downward \citep[eg.,][]{TH92}.  Summing all the
molecular gas (Figure 5c) over the central $\sim$90 pc gives
M$_{H_{2}}\rm ~\simeq~ 5.3 \times 10^{6}~ M_{\odot}$ (corrected for He
and ``missing flux'').  From the nuclear rotation curve, the estimated
dynamical mass over the same region is M$_{dyn}~\simeq~ 5.7 \times
10^{7}\rm~ M_{\odot}$ \citep{TH92}, or M$_{H_{2}}$/M$_{dyn}~ \simeq ~
10\%$.  The molecular mass fraction increases to
M$_{H_{2}}$/M$_{dyn}~\simeq~30\%$ within the central 45 pc.  If
$X_{CO}$ drops substantially outside of the central starburst region
as indicated by the C$^{18}$O data, then the molecular arms do not
make up a substantial percentage of the dynamical mass over the
central arcminute.

The dynamics (and hence the molecular gas distribution) of the nuclear
region are dominated by the stellar mass distribution.  Recent NIR
K-band images indicate that a small stellar bar (major axis $\sim$110
pc) is present in the nucleus \citep{BFG97}.  The molecular gas
distribution probably reflects gas moving on barred orbits
\citep*{Loet84, I90, BFG97}.  Using the $^{13}$CO observations,
\citet{TH92} estimated that star formation efficiencies (SFE) towards
the central starburst region and GMC B, in particular, are very high.
They estimate based on N$_{Lyc}$ that, at the starburst, the total
mass of newly formed stars with masses greater than 1 M$_{\odot}$,
M$_{*}$, is $\simeq~10^{6}~M_{\odot}$.  With our improved H$_{2}$ mass
for GMC B of M$_{H_{2}}~\simeq~ 4.0 \times 10^{5}\rm~ M_{\odot}$, we
obtain SFE of M$_{*}$/M$_{H_{2}}~\sim$2.  This is an extremely high
SFE.  Alternatively, this could be an indication that the IMF is
truncated above 1 M$_{\odot}$ or that the starburst is in the process
of destroying its molecular cloud.

It is worth reemphasizing a point made by other studies that the sites
of star formation in IC 342 are not strictly correlated with either
the global molecular gas column density or molecular gas density
\citep{I90, D92, WITHL93}.  Active star formation is seen only in two
of the five central molecular clouds.  The molecular cloud densities
and temperatures obtained from the CO LVG analysis and HCN emission do
not differ by more than a factor of $\sim$2 across the five GMCs, but
the star formation varies substantially \citep[\S 3.2.3;][]{D92}.  The
locations of star formation appear to be determined predominantly by
dynamics.  The two major sites of star formation are located roughly
perpendicular to the major axis of the small scale NIR bar, at the
intersection with the spiral arms.  This orientation suggests that
these star formation sites are at the intersection of the so-called
$x_{1}$ and $x_{2}$ orbits \citep[eg.,][] {RHV79, A92, HS96}.  

From CO bandhead spectroscopy, \citet{BFG97, BVV99} find that the age
of the central cluster located in the Central Trough (not to be
confused with the starburst) to be 10$^{6.8-7.8}$ years old.  Based on
chemical enrichment models, similar ages are required to explain the
anomalous C$^{18}$O abundances we find in this region \citep{HM93}.
However, the sites of current star formation (GMCs B \& C) are
probably younger because they are still sites of radio detected HII
regions.  The high emission measures \citep[EM $\gapprox ~ 10^{5}~\rm
cm^{-6}pc$;][]{TH83} and the large implied number of O stars suggest
that the starburst is younger than 10 Myr, perhaps significantly, and
hence younger than the central cluster.  Star formation along the arms
of the mini-spiral farther away from the nucleus also appears
dynamically induced, perhaps by shocks associated with the bar orbits.
The star formation along the leading edge of the spiral arms seen in
H$\alpha$ emission can be attributed to the arm molecular gas which is
currently upstream \citep{TH92}.  Taking into account the rotation
speed of material in the arms, it is estimated that the star formation
is as little as 1 Myr ``downstream'' from the molecular spiral arms
\citep{TH92}.  Therefore, the star formation on the leading edges of
the arms are probably quite young as well (and presumably roughly
coeval with the starburst).  These result provide additional evidence
for the following basic scenario for the recent star formation history
of IC 342's nucleus \citep{BFG97}.  Apparently a star formation event
$\sim$10 - 30 Myr ago formed the stars in the central cluster/NIR bar.
The bar potential drove the remaining molecular gas into the nucleus
along the $x_{1}$-orbits traced by the CO arms.  The gas then collided
with molecular gas on the perpendicular $x_{2}$-orbits, compressing
the gas and triggering the younger currently seen starburst.  Such
dynamically induced star formation may then explain why such high SFEs
seen only in the central two intersection regions and not elsewhere.

\section{Summary}

Aperture synthesis maps of C$^{}$O(1-0) and (2-1) in the nucleus of IC
342 are presented.  This completes an extensive program of mapping the
J=2-1 and J=1-0 transitions of CO and its isotopomers.  The H$_{2}$
column density as traced by optically thin C$^{18}$O is similar in
morphology to what is implied from $^{12}$CO observations, except for
two main differences.  Firstly, the arms of the nuclear mini-spiral
seen in C$^{18}$O (and $^{13}$CO) emission are more symmetric than
those seen for $^{12}$CO.  Isotopic observations demonstrate that the
asymmetry in brightness of the $^{12}$CO arms are not due to large
differences in amount of molecular gas but instead due to asymmetries
in optically thick $^{12}$CO emissivity.  The northern arm is viewed
with the starburst ``lit'' face towards us, appearing brighter in
optically thick $^{12}$CO, while for southern arm is viewed from the
back, with the illuminated side away from the observer.

Secondly, C$^{18}$O is weaker than expected from the $^{12}$CO maps
across the starburst region.  If C$^{18}$O accurately traces the
amount of molecular gas present, the standard conversion factor from
$^{12}$CO to H$_{2}$, $X_{CO}$, must be lower than the Galactic value.
Comparisons between the column density measured from C$^{18}$O, from
1.3 mm dust continuum emission, and from the virial theorem imply that
$X_{CO}(IC ~342)$ is about 1/2 - 1/3.  The Galactic $X_{CO}$
overestimates the amount of H$_{2}$ gas in IC 342.  For the diffuse
gas component at larger distances from the center of IC 342, the
conversion factor is about a factor of 2 - 3 lower, yet.  As a result,
a simple scaling of the Galactic conversion factor to an appropriate
value for IC 342 is not good enough to characterize the true column
density across the starburst.

The mean H$_{2}$ column density for the central GMCs is
$<N_{H_{2}}>~\simeq ~4 - 5\times 10^{22}~\rm cm^{-2}$ ($A_{v}~\sim 40
- 50$).  This corresponds to cloud masses of $\sim ~0.5 - 1 \times
10^{6}~\rm M_{\odot}$.  Excitation temperatures derived from the
C$^{18}$O data are T$_{ex}~\sim ~ 7~-~10$ K over much of the central
100 pc.  This value is low compared with what is found from the
optically thicker $^{12}$CO ($\gapprox 30$ K) and $^{13}$CO (20 K)
transitions.  C$^{18}$O must be subthermally excited over the entire
range, indicating that the average densities of the GMCs are not
higher than $n_{crit}$(C$^{18}$O(2-1)) (or $\simeq 10^{4}~\rm
cm^{-3}$).  Densities and temperatures obtained for the GMCs from an
LVG analysis of $^{13}$CO and C$^{18}$O are $<n_{H_{2}}>~\simeq ~
10^{3.5}~\rm cm^{-3}$ and T$_{kin} ~\simeq ~ 10-40$ K.  Average
densities estimated from the total column densities are consistent
with these values.  These values are high relative to Galactic disk
GMCs, but low compared to what is seen is other nearby starburst
galaxies \citep[eg.,][]{JPCR95}.

Our new results indicate that star formation efficiencies are
extremely high at the starburst, with $M_{*}/M_{gas}~\sim ~ 2$ over
the central $\sim$45 pc.  These measurements also indicate that
molecular mass fractions in the center of IC 342 are lower than
previously estimated, with $M_{gas}/M_{dyn}$ $\sim$10 \% for the inner
90 pc rising to $\sim$30 \% over the central $\sim$50 pc.The locations
of star formation in the central $\sim$200 pc of IC 342 appear to be
determined by large scale dynamical forces rather than gas surface
density.  Star formation occurs primarily at the ends of the
mini-spiral arms, oriented perpendicular to the large scale ``bar''.
This may be the location of the $x_{1}~-~x_{2}$ orbit crossings.
Weaker star formation is seen downstream of the spiral arms apparently
triggered by there passage.

Very low $^{12}$CO(1-0)/C$^{18}$O(1-0) and
$^{13}$CO(1-0)/C$^{18}$O(1-0) line ratios seen towards the optically
thin Central Trough region provide evidence that the C$^{18}$O
abundance is enhanced due to ejecta from the massive stars of the
central starburst.

\acknowledgements

We are grateful to John Carpenter, Dave Frayer, Charlie Qi, Steve
Scott and Dave Woody for their support and assistance in making the
observations at OVRO.  We also thank Mike Jura for reading the
manuscript before submission.  We thank the referee for comments that
improved the paper.  This work was supported in part by NSF grant
AST-0071276.  The Owens Valley Millimeter Interferometer is operated
by Caltech with the support from the NSF.

\clearpage

\figcaption{C$^{18}$O(1-0) channel map.  LSR velocities
are listed at the top right of each plane.  The contours are plotted
in intervals of 12 mJy bm$^{-1}$ (or 0.12 K for the 3.$^{''}$3 x
3.$^{''}$0 beam) corresponding to 2$\sigma$.  The beam is plotted in
the bottom left of the first plane.}

\figcaption{C$^{18}$O(2-1) channel map.  LSR velocities
are listed on each plane and are the same as in Figure 1.  The
contours are plotted in intervals of 32 mJy bm$^{-1}$ (or 0.29 K for
the 1.$^{''}$8 x 1.$^{''}$6 beam) corresponding to 2$\sigma$.  The
beam is plotted in the bottom left of the first plane.}

\figcaption{The integrated intensity of the transitions. (a)
$^{12}$CO(1-0) integrated intensity (Levine et al. 1994).  The
$^{12}$CO(1-0) map has been convolved to the beamsize of the
C$^{18}$O(1-0) map.  The contours are in steps of 47.1 K km
s$^{-1}$. (b) C$^{18}$O(1-0) integrated intensity.  Contours are in
steps of 1.32 K km s$^{-1}$ ($2\sigma$), minus the $2\sigma$ contour.
(c) C$^{18}$O(2-1) integrated intensity.  Contours are the same as (b)
with a $2\sigma$ value of 2.68 K km s$^{-1}$.  The large circle marks
the FWHM power points of the mosaicked (2-1) field.  The positions of
the fitted GMCs are labeled on plot.}

\figcaption{The total H$_{2}$ column density maps as
derived from C$^{18}$O (see text) overlaid on the HST H$\alpha$ image
of IC 342 (Gallagher et al, in preparation).  The contours are in steps
of $\rm 6.0\times 10^{21}~cm^{-2}$.}

\figcaption{Gas excitation maps.  Locations with integrated
intensities below 3$\sigma$ in either of the generating maps are
blanked. (a) The C$^{18}$O(2-1)/C$^{18}$O(1-0) line ratio map.  The
greyscale range is 0 - 1.5, with a peak value of 2.2 at the location
of the starburst. (b) T$_{ex}$ derived from the
C$^{18}$O(2-1)/C$^{18}$O(1-0) ratio (\S 3.3.1).  Greyscale range is 1
- 15 K, with a peak of 19 K.  As a reference, the $^{12}$CO(1-0)
integrated intensity map (contoured in steps of 125 K km s$^{-1}$) is
overlaid on each ratio map.  (c) The molecular gas column density
derived from the C$^{18}$O(1-0) intensity map (Figure 3b), the
T$_{ex}$ map in (b), and a [H$_{2}$/C$^{18}$O] abundance ratio of
$2.94\times 10^{6}$.  The contours are in steps of $2.5 \times
10^{20}\rm ~cm^{-2}$, starting with $7.5 \times 10^{20}\rm ~cm^{-2}$.}

\figcaption{CO isotopic line ratio maps.  Locations with
integrated intensities below 3$\sigma$ in either of the generating
maps are blanked. (a) The $^{12}$CO(1-0)/$^{13}$CO(1-0) ratio map
(Wright et al 1993; MTH00).  The greyscale ranges from 4 - 12 with
darker shades corresponding to higher ratios.  (b)
$^{12}$CO(1-0)/C$^{18}$O(1-0) line ratio map, with the greyscale
ranging from 0 - 75. (c) The $^{13}$CO(1-0)/C$^{18}$O(1-0) line ratio
map, with a greyscale range of 0 - 10.  The $^{12}$CO(1-0) contours of
Figure 6 are shown for reference.  For maps that include the
$^{13}$CO(1-0) transition, the beamsize is 4.$^{''}$8 x 3.$^{''}$9
(MTH00), while the others are those of C$^{18}$O(1-0).}

\figcaption{Large Velocity Gradient model solutions
for six nuclear locations.  Contoured region represents the 1$\sigma$
confidence solution for the observed antenna temperature and
(2-1)/(1-0) ratio.  Solutions are displayed for both $^{13}$CO (black
line; MTH00) and C$^{18}$O (grey line), with two assumed values of
$X_{CO}$/$dv/dr$.  Abundance ratios of [$^{13}$CO/H$_{2}$] =
$2.1\times 10^{-6}$ and [C$^{18}$O/H$_{2}$] = $3.4\times 10^{-7}$ are
assumed.  Solid lines are the solutions for
$dv/dr~=~1~km~s^{-1}~pc^{-1}$ ($X_{13CO}/dv/dr ~=~10^{-5.68}$;
$X_{C18O}/dv/dr~=~10^{-6.47}$), consistent with the data (Table 4).
Dashed lines are solutions for $dv/dr~=~5~km~s^{-1}~pc^{-1}$
($X_{13CO}/dv/dr ~=~10^{-6.38}$; $X_{C18O}/dv/dr~=~10^{-7.17}$).
Solutions are based on $4.^{''}8 \times 3.^{''}9$ resolution and a
filling factor of 0.25 for both $^{13}$CO and C$^{18}$O (MTH00; Table
4).  A decrease of a factor of two in filling factors will shift the
solution up ($\sim$0.4 dex) and to the left ($\sim$ 10 K).  Dot-dashed
lines show the beam-averaged density estimated via the virial theorem,
for each GMC (using the full C$^{18}$O(2-1) resolution data; Table
4).}

\figcaption{2.6 mm and 1.3 mm continuum maps. (a) 2.6 mm (110
GHz) continuum emission.  The contours are 2$^{0.5i}$ (i=0,1,2...)
times the 2$\sigma$ value of 0.4 mJy bm$^{-1}$. (b) 1.3 mm (220 GHz)
continuum emission.  The contours levels are the same as 2.6 mm, with a
2$\sigma$ value of 4.0 mJy bm$^{-1}$.  Both maps are displayed of
the same resolution (3 mm). (c) The 2 cm/ 2.6 mm spectral index map
($S\propto \nu ^{\alpha}$).  (d) The 2.6 mm/ 1mm spectral index map.
For the two spectral index maps darker greyscales represent larger
values.  Contours are in steps of 0.25 for (c) and 0.5 for
(d). Negative contours are dashed.  The maps are blanked when 2.6 mm
continuum is less than 0.4 mJy bm$^{-1}$ (c) and when 1.3 mm continuum
is less 2.0 mJy bm$^{-1}$ (d).}

\figcaption{The ratio of the conversion factor found from
C$^{18}$O observations in IC 342 versus the Galactic value as a
function of distance from the dynamical center of IC 342 (Table 1).
The error bars only reflect the uncertainty in the C$^{18}$O
integrated intensity, and does not include systematic errors such as
those associated with errors in derived T$_{ex}$,
uncertainties in the [C$^{18}$O/H$_{2}$] abundance ratio or errors in
the Galactic $X_{CO}$.}

\clearpage
\begin{deluxetable}{lcc}
\tablenum{1}
\tablewidth{0pt}
\tablecaption{IC 342  Basic Data}
\tablehead{\colhead{Characteristic} 	& \colhead{Value}	&
\colhead{Reference}}
\startdata
Revised Hubble Class      &Sab(rs)cd   & 1  \\
Dynamical Center        & $\alpha(2000) = 03^{h} 46^{m} 48^{s}.7\pm 0.^{s}3
~~~$ &2   \\
  &$ \delta(2000) = +68^{o} 05' 46.^{''}8\pm 2^{''}$   &   \\
2$\mu$m peak$~~~~$     & $\alpha(2000) = 03^{h} 46^{m} 48^{s}.3\pm 0.^{s}3$
  &3   \\
  &$ \delta(2000) = +68^{o} 05' 46''.8\pm 2^{''}$       &   \\
$\ell^{II}$,b$^{II}$ & 138.2$^{o}$,$+10.6^{o}$  &1   \\
V$_{lsr}$   &35 kms$^{-1}$   &2   \\
Adopted Distance  & 1.8 Mpc  &4   \\
Inclination  & 30$^{o}$  &5   \\
Position Angle  &39.4$^{o}$   &5   \\
B$_{T}$  & 9.16 mag  &1   \\
\tablecomments{Units of right ascension are hours, minutes and seconds, and 
units of declination are degrees, arcminutes, and arcseconds.}
\tablerefs{(1) de Vaucouleurs, de Vaucouleurs \& Corwin 1976; (2)
TH92; (3) Becklin et al. 1980; (4) McCall 1989; 
(5) Crosthwaite, Turner \& Ho 2000} 
\enddata
\end{deluxetable}
\clearpage
\begin{deluxetable}{lcccccc}
\tablenum{2}
\tablewidth{0pt}
\tablecaption{Observational Data\tablenotemark{a}}
\tablehead{
\colhead{Transition\tablenotemark{b}} 
&\colhead{Frequency} 
& \colhead{T$_{ssb}$} 
&\colhead{$\Delta V_{chan}$}
&\colhead{$\Delta \nu_{band}$} 
& \colhead{Beamsize} 
& \colhead{Noise level} \\
\colhead{} 
&\colhead{\it (GHz)} 
& \colhead{(K)} 
&\colhead{($km~s^{-1}$)}
&\colhead{\it (MHz)}
& \colhead{\it (arcsec; deg)} 
& \colhead{\it (K / mJy Bm$^{-1}$)}} 
\startdata
C$^{18}$O(1-0)& 109.782 & 400-450 & 10.92& 128 
&$3.3\times 3.0;-43^{o}$ & 0.062/6\\
C$^{18}$O(2-1)& 219.560 &430-1050 &5.46 &128 
&$1.8\times 1.6;-26^{o}$& 0.14/16\\
3mm Cont. & 109.5   &360-430 &\nodata  &1000 
&$3.3\times 3.0;-44^{o}$ & 0.004 /0.40\\
1mm Cont. & 219.8   &400-1000 &\nodata  &1000 
&$1.9\times 1.6;-15^{o}$ & 0.016/2\\

\tablenotetext{a}{For observations made from 1997 October 22 - 1997 
December 12}
\tablenotetext{b}{Phase Center \#1:  V$_{LSR}$=35 $km ~s^{-1}$ 
$\alpha = 03^{h} 46^{m} 48^{s}.10~~\delta = +68^{o} 05' 44.^{''}8$ (J2000)\\
\#2: $\alpha = 03^{h} 46^{m} 49^{s}.01~~\delta = +68^{o} 05' 47.^{''}8$ 
(J2000)}
\enddata
\end{deluxetable}

\clearpage

\begin{deluxetable}{lcccccc}
\tablenum{3} 
\tablewidth{0pt} 
\tablecaption{Measured Intensities \& Ratios\tablenotemark{a}} 
\tablehead{ 
\colhead{Location} 
&\colhead{Position}
& \colhead{I(C$^{18}$(1-0))} 
& \colhead{I(C$^{18}$(1-0))} 
& \colhead{$\frac{C^{18}O(2-1)}{C^{18}O(1-0)}$\tablenotemark{b}} 
& \colhead{$\frac{^{12}CO(1-0)}{C^{18}O(1-0)}$\tablenotemark{b}}
& \colhead{$\frac{^{13}CO(1-0)}{C^{18}O(1-0)}$\tablenotemark{b}}
\\
\colhead{}
& \colhead{($3^{h}46^{m}$, $68^{0} 05'$)}
& \colhead{(K kms$^{-1}$)}  
& \colhead{(K kms$^{-1}$)}
& \colhead{}
& \colhead{}
& \colhead{}}
\startdata
A & 48.$^{s}$51,43.''4&8.7$\pm$1.3 & 26$\pm$6.4 
& 1.7$\pm$0.50 & 67$\pm$14 & 5.8$\pm$1.2 \\
B & 47.$^{s}$78,46.''0 & 7.0$\pm$1.3 & 25$\pm$6.2 
& 2.2$\pm$0.68 & 69$\pm$16 & 7.2$\pm$1.7 \\
C & 49.$^{s}$19,49.''5 &22$\pm$3.3 & 43$\pm$10 
& 1.2$\pm$0.34 & 32$\pm$7 & 5.4$\pm$1.2 \\
D & 48.$^{s}$93,59.''2 & \mbox{$\stackrel{<}{_{\sim}}$}4.0 & 16$\pm$4 
& $\sim$1.0 & \mbox{$\stackrel{>}{_{\sim}}$}100 
& \mbox{$\stackrel{>}{_{\sim}}$}10 \\
E & 47.$^{s}$50,42.''0 & 21$\pm$3.2 & 26$\pm$6.4 & 0.63$\pm$0.19
& 28$\pm$6 & 4.1$\pm$0.90  \\
Arm\tablenotemark{c} &47.$^{s}$10,33.''9 & 4.0$\pm$1.3 &\nodata 
& \nodata & $\sim$51 & $\sim$4.7 \\
Off-Arm\tablenotemark{c}&48.$^{s}$10,03.''8\tablenotemark{d} 
& \mbox{$\stackrel{<}{_{\sim}}$}4.0 &\nodata & \nodata 
& \mbox{$\stackrel{>}{_{\sim}}$}110 
& \mbox{$\stackrel{>}{_{\sim}}$}7 \\
C. Trough\tablenotemark{c} &48.$^{s}$26,47.''8  &5.3$\pm$1.3 
& $<$8.0 & \nodata & 24$\pm$7 & 3.2$\pm$1.0 \\
\tablenotetext{a}{The measurements of the intensities are based on the
resolutions given in Table 2, while the resolutions of the ratio maps
are those of the C$^{18}$O(1-0) (expect the
$^{13}$CO(1-0)/C$^{18}$O(1-0) ratio which has a resolution of
$4.^{''}8\times 3.^{''}9$).  The uncertainties reflect the larger of 
the absolute calibration uncertainty or the map noise.}
\tablenotetext{b}{Ratios are for integrated intensity, $\int T_{b}dv$, 
rather than the peak intensities.}
\tablenotetext{c}{The positions are taken from MTH00.}  
\tablenotetext{d}{Based on $\alpha = 3^{h}46^{m}~ \delta = 68^{0}06'~$} 
\enddata
\end{deluxetable}

\clearpage

\begin{deluxetable}{lcccccc}
\tablenum{4}
\tablewidth{0pt}
\tablecaption{Giant Molecular Clouds in IC 342}
\tablehead{
\colhead{Cloud} 
&\colhead{$\alpha_{o}, ~\delta_{o}$}
&\colhead{$a \times b;~ pa$} 
&\colhead{$\Delta v_{1/2}$}
&\colhead{$M_{vir}$}
&\colhead{$<^{vir}N_{H_{2}}>$\tablenotemark{b}}
&\colhead{$M_{C18O}$\tablenotemark{c}}
\\
\colhead{} 
&\colhead{(03 46;68 05)} 
&\colhead{($pc \times pc;^{o}$)} 
&\colhead{($km~s^{-1}$)}
&\colhead{($10^{6}~M_{\odot}$)}
&\colhead{($10^{3}~cm^{-3}$)}
&\colhead{($10^{6}~M_{\odot}$)}}
\startdata
A &48.51;43.4&$29\times 10;~78^{o}$&25&1.4 & 5.8&0.48 \\
B1&47.80;46.5&$18\times 14;~0.7^{o}$&15&0.46 & 2.3& 0.94\\
B2&47.77;45.5&$29\times 7.8;~30^{o}$&21&0.86 & 5.2& \\
C1&49.15;51.3&$22\times 16;~17^{o}$&27&1.8 & 5.6& 1.2\\
C2&49.19;49.5&$39\times 8.7;~180^{o}$&26&1.6 & 5.2& \\
C3&49.16;46.9&$37\times 0;~20^{o}$\tablenotemark{a}&29&$<$1.7 & 
$<$9.3& \\
E1&47.53;42.7&$28\times 12;~34^{o}$&19&0.90 & 3.0& 1.0\\
E2&47.45;41.2&$19\times 16;~140^{o}$&15&0.54 & 2.1&  \\
\enddata
\tablenotetext{a}{The cloud is considered unresolved since the Gaussian fit 
size is less than half of the beam minor axis of $1.^{''}5$ (13 pc).}
\tablenotetext{b}{The average density assumes that the cloud is an 
ellipsoid of line of sight dimension, $\ell \simeq 1.4 \sqrt{ab}$}
\tablenotetext{c}{Values represent the masses for the whole GMCs, since 
substructure is not resolved in C$^{18}$O(1-0).}
\end{deluxetable}
\clearpage

\begin{deluxetable}{lccccc}
\tablenum{5}
\tablewidth{0pt}
\tablecaption{The Conversion Factor in IC 342}
\tablehead{
\colhead{Location} 
&\colhead{$^{C^{18}O}N_{H_{2}}$\tablenotemark{a}} 
&\colhead{$^{vir}N_{H_{2}}$\tablenotemark{b}}
&\colhead{$^{dust}N_{H_{2}}$\tablenotemark{c}}
&\colhead{$^{X_{CO}}N_{H_{2}}$\tablenotemark{d}}
&\colhead{$\left<\frac{X_{MW}}{X_{C18O}}\right>$}
\\
\colhead{} 
&\colhead{($10^{22}~cm^{-2}$)} 
&\colhead{($10^{22}~cm^{-2}$)} 
&\colhead{($10^{22}~cm^{-2}$)} 
&\colhead{($10^{22}~cm^{-2}$)} 
&\colhead{}}
\startdata
A&3.1&11&$\sim$1.5&15&4.8 \\
B&3.3&11&3.8&11&3.3 \\
C&4.8&13\tablenotemark{f}&7.7&16&3.3 \\
D&1.3&\nodata&\nodata&13&10 \\
E&3.7&12&\nodata&15&4.1 \\
Arm&1.1\tablenotemark{e}&\nodata&\nodata&7.6&6.9 \\
Off Arm&$<$0.55\tablenotemark{e}&\nodata&\nodata&3.7&$>$6.7 \\
C. Trough&1.1\tablenotemark{e}&\nodata&\nodata&5.1&4.6 \\ 
\enddata
\tablenotetext{a}{Using an assumed abundance of [C$^{16}$O/C$^{18}$O] = 
250, and correcting for the 67\% missing flux.}
\tablenotetext{b}{Based on the sum of the masses in each subcomponent 
within the $^{12}$CO beamsize divided by 1.36 $\mu_{H_{2}}$ (to 
account for He) and averaged over the beam.}
\tablenotetext{c}{Based on the dust mass, $M_{H_{2}}(M_{\odot})$ 
estimate averaged over the beam (\S 3.4).}
\tablenotetext{d}{Using a $X_{CO}(MW) = 2 \times 10^{20} 
~cm^{-2} K~km~s^{-1}$, and the $^{12}$CO(1-0) map in Figure 3a, 
corrected for the 25\% missing flux.}
\tablenotetext{e}{Assuming an excitation temperature of 10 K.}
\tablenotetext{f}{Only component C2 is included, C1 and C3 are outside 
the $^{12}$CO(1-0) beamsize.}
\end{deluxetable}

\end{document}